\documentclass[aps,prl,twocolumn,superscriptaddress,nofootinbib,preprintnumbers]{revtex4-1}
\usepackage{amsmath}
\usepackage{graphicx}
\usepackage{subfigure}
\usepackage{amssymb}
\usepackage{xcolor}
\usepackage{multirow}
\usepackage{cancel}
\usepackage{color}
\usepackage{ulem}
\usepackage{listings}
\usepackage{xcolor}
\usepackage{float}
\usepackage{slashed}

\usepackage{wrapfig}
\usepackage{mathtools}

\usepackage{tikz}
\usetikzlibrary{arrows,shapes}
\usetikzlibrary{trees}
\usetikzlibrary{matrix,arrows} 				
\usetikzlibrary{positioning}				
\usetikzlibrary{calc,through}				
\usetikzlibrary{decorations.pathreplacing}  
\usepackage{pgffor}							

\usetikzlibrary{decorations.pathmorphing}	
\usetikzlibrary{decorations.markings}
\tikzset{
    vector/.style={decorate, decoration={snake}, draw},
	provector/.style={decorate, decoration={snake,amplitude=2.5pt}, draw},
	antivector/.style={decorate, decoration={snake,amplitude=-2.5pt}, draw},
    fermion/.style={draw=black, postaction={decorate},
        decoration={markings,mark=at position .55 with {\arrow[draw=black]{>}}}},
    fermionbar/.style={draw=black, postaction={decorate},
        decoration={markings,mark=at position .55 with {\arrow[draw=black]{<}}}},
    fermionnoarrow/.style={draw=black},
    gluon/.style={decorate, draw=black,
        decoration={coil,amplitude=4pt, segment length=5pt}},
    scalar/.style={dashed,draw=black, postaction={decorate},
        decoration={markings,mark=at position .55 with {\arrow[draw=black]{>}}}},
    scalarbar/.style={dashed,draw=black, postaction={decorate},
        decoration={markings,mark=at position .55 with {\arrow[draw=black]{<}}}},
    scalarnoarrow/.style={dashed,draw=black},
    electron/.style={draw=black, postaction={decorate},
        decoration={markings,mark=at position .55 with {\arrow[draw=black]{>}}}},
	bigvector/.style={decorate, decoration={snake,amplitude=4pt}, draw},
}

\tikzstyle{block} = [draw, rectangle, 
    minimum height=3em, minimum width=6em]

\usepackage[colorlinks,citecolor=blue]{hyperref}
\usepackage{amsmath}
\usepackage{wrapfig}

\usepackage[T1]{fontenc} 
\usepackage[utf8]{inputenc} 
\usepackage{times}

\newcommand{\be}{\begin{equation}}
\newcommand{\ee}{\end{equation}}
\newcommand{\beq}{\begin{equation}}
\newcommand{\eeq}{\end{equation}}
\newcommand{\bea}{\begin{eqnarray}}
\newcommand{\eea}{\end{eqnarray}}
\newcommand{\besp}{\begin{equation}\begin{split}}
\newcommand{\eesp}{\end{split}\end{equation}}

\newcommand{\nn}{\nonumber}

\newcommand{\Eq}[1]{Eq.~(\ref{#1})}

\newcommand{\Dfbd}{\mathord{\buildrel{\lower3pt\hbox{$\scriptscriptstyle\leftrightarrow$}}\over {D}_{\mu}}}

\def\mB{\mathcal{B}}

\def\mD{\mathcal{D}}
\def\mE{\mathcal{E}}

\def\mI{\mathcal{I}}

\def\mL{\mathcal{L}}

\def\mO{\mathcal{O}}
\def\mP{\mathcal{P}}
\def\mQ{\mathcal{Q}}

\def\mV{\mathcal{V}}

\def\0{\textbf{0}}
\def\1{\textbf{1}}
\def\2{\textbf{2}}
\def\3{\textbf{3}}
\def\4{\textbf{4}}
\def\5{\textbf{5}}
\def\6{\textbf{6}}
\def\7{\textbf{7}}
\def\8{\textbf{8}}
\def\9{\textbf{9}}

\def\v{\textbf{v}}
\def\d{\text{d}}

\usepackage{fontawesome5}

\definecolor{RoyalBlue}{cmyk}{1, 0.50, 0, 0}

\begin{document}

\title{Reviving primordial black hole formation in slow first-order phase transitions}

\author{Wen-Yuan Ai}
\email{wenyuanai@sjtu.edu.cn}
\affiliation{State Key Laboratory of Dark Matter Physics,
Tsung-Dao Lee Institute and School of Physics and Astronomy, Shanghai Jiao Tong University, Shanghai 201210, China}
\affiliation{Key Laboratory for Particle Astrophysics and Cosmology (MOE), and Shanghai Key Laboratory for Particle Physics and Cosmology, Shanghai Jiao Tong University, Shanghai 201210, China}

\author{Ke-Pan Xie}
\email{kpxie@buaa.edu.cn, corresponding author.}
\affiliation{School of Physics, Beihang University, Beijing 100191, China}

\begin{abstract}
Large curvature perturbations generated during slow first-order phase transitions are a promising source of primordial black holes. However, recent analyses suggested that the mechanism is ruled out once the density contrast and the formation threshold are evaluated in the same gauge. In this work, we show that this mechanism remains viable: after a supercooled transition, reheating can be sufficiently slow that the Universe enters an early matter-dominated era, during which even small overdensities grow and collapse into primordial black holes.
\end{abstract}
\maketitle

{\it Introduction}---Spontaneous symmetry breaking in the early Universe may proceed via a first-order phase transition (FOPT), generating gravitational waves (GWs) that could be detectable today~\cite{Mazumdar:2018dfl,Caprini:2019egz,Athron:2023xlk}. In a supercooled transition, the Universe typically undergoes a period dominated by vacuum energy before entering a phase dominantly governed by other components, most commonly radiation. Owing to the stochastic nature of bubble nucleation, the transition time varies across different spatial regions, such that some regions start transiting later than others. Since the vacuum energy density remains constant while radiation dilutes due to cosmic expansion, these delayed-transition regions develop an overdensity relative to the surrounding plasma~\cite{Liu:2021svg}. Primordial black holes (PBHs) can form if the density contrast $\delta = (\rho - \bar{\rho}) / \bar{\rho}$ exceeds a critical threshold $\delta_c \approx 0.45$~\cite{Musco:2004ak,Harada:2013epa}. This mechanism has been extensively studied~\cite{Kodama:1982sf,Hall:1989hr,Liu:2021svg,Hashino:2021qoq,Hashino:2022tcs,He:2022amv,Kawana:2022olo,Lewicki:2023ioy,Gouttenoire:2023naa,Gouttenoire:2023pxh,Jinno:2023vnr,Banerjee:2023brn,Banerjee:2023qya,Baldes:2023rqv,Salvio:2023blb,Conaci:2024tlc,Lewicki:2024ghw,Flores:2024lng,Kanemura:2024pae,Cai:2024nln,Goncalves:2024vkj,Banerjee:2024fam,Arteaga:2024vde,Banerjee:2024cwv,Wu:2024lrp,Hashino:2025fse,Zhang:2025kbu,Huang:2025hos,Cao:2025jwb,Kierkla:2025vwp,Huang:2025hos,Wang:2025hwc,Ning:2026nfs}.

Among these studies, a particularly influential framework was developed in Ref.~\cite{Lewicki:2024ghw}, where the bubble nucleation history in supercooled FOPTs is simulated to obtain the probability distribution function (PDF) $P(\delta)$, along with semi-analytic estimates of the PBH mass and abundance. However, a critical issue has recently been pointed out~\cite{Franciolini:2025ztf}: the density contrast $\delta$ is gauge-dependent in relativistic perturbation theory. The quantity extracted from bubble simulations corresponds to the spatially flat gauge, $\delta^{(F)}$, whereas the conventionally cited value of PBH formation threshold is defined in the comoving gauge, $\delta^{(C)}$. These two quantities are related at horizon re-entry roughly by $\delta^{(F)} \simeq 10\,\delta^{(C)}$, which effectively suppresses PBH formation to negligible levels. A consistent conclusion was independently reached in Ref.~\cite{Wang:2026zvz} based on a gauge-invariant analysis employing approximate superhorizon relations.

In this Letter, we revisit this problem using a new method to compute $\delta^{(C)}$ in supercooled FOPTs. Our approach is based on gauge-invariant observables and does not rely on the superhorizon approximation. We first reproduce the results of Refs.~\cite{Franciolini:2025ztf,Wang:2026zvz}, confirming the gauge-dependence issue. Crucially, however, we then show that PBH formation {\it remains viable}: the post-FOPT reheating stage may be inefficient, leading to a transient early matter-dominated (EMD) era during which small density perturbations grow rapidly and collapse into PBHs. A model-independent analysis reveals that $\beta/H_n\lesssim18$ together with an EMD duration $\gtrsim10^3 H_*^{-1}$ yields abundant PBH production, and we demonstrate that these conditions are naturally realized in a broad range of new-physics models.

{\it Vacuum energy perturbation}---The bubble nucleation rate can be parameterized as
\be
\Gamma(t)=H_n^4\,e^{\beta (t-t_n)}\,,
\ee
where $H_n$ is the Hubble rate at nucleation time $t_n$, and $\beta^{-1}$ sets the FOPT duration. With $a(t)$ the scale factor and $H(t)=\dot a(t)/a(t)$ the Hubble rate, in a supercooled FOPT, the comoving Hubble radius $(aH)^{-1}$ first shrinks during vacuum domination and later expands after the vacuum energy is released. Its minimum defines percolation moment $t_*$. Therefore, the largest comoving scale that exits the horizon is $k_{\rm max}=a_*H_*$ with $a_*\equiv a(t_*)$ and $H_*\equiv H(t_*)$. After percolation, any mode with $k<k_{\rm max}$ re-enters the horizon at a later time $t_k$ given by $k=a_kH_k$ with $a_k\equiv a(t_k)$ and $H_k\equiv H(t_k)$.

The average false vacuum volume fraction is~\cite{Guth:1981uk}
\be\label{barFt}
\bar{F}(t)=\exp{\left[-\frac{4\pi}{3}\int_{-\infty}^t \d t'\, \Gamma(t')\left(\int_{t'}^t \d t''\frac{a(t')}{a(t'')}\right)^3\right]}\,,
\ee
where we take the bubble wall velocity $v_w\approx1$, as appropriate for a supercooled transition. The background vacuum energy density then evolves as $\bar\rho_V(t)=\bar{F}(t)\Delta V$, with $\Delta V$ the energy density difference between the false and true vacua. Due to the randomness of bubble nucleation, a specific spherical comoving region of volume $4\pi k^{-3}/3$ has a false-vacuum fraction $F_k(t)$ that fluctuates around $\bar F(t)$. This induces a vacuum energy perturbation $\delta\rho_V=(F_k-\bar F)\Delta V$, which sources the density contrast $\delta$.

To determine $\delta$ from $\delta \rho_V$, it is necessary to specify the underlying energy transfer process. In a supercooled FOPT, the scalar field $\phi$ first tunnels through the potential barrier to a field value that may still lie far from the true vacuum. It subsequently rolls down toward the true vacuum, undergoes coherent oscillations about the minimum, and eventually decays. If its decay width satisfies $\Gamma_\phi \gtrsim H_*$, reheating is fast, and vacuum energy converts directly into radiation. If instead $\Gamma_\phi < H_*$, reheating is slow and the oscillating $\phi$ behaves as nonrelativistic matter, causing an EMD era. The full dynamics thus factorizes into two successive energy transfers: from vacuum to a matter-like scalar field, and subsequently from scalar to radiation. To treat both fast and slow reheating scenarios in a unified framework, we model the Universe during the FOPT as a two-fluid system: a vacuum component with equation of state $P_V = -\rho_V$, and a second component $f$ characterized by $P_f = w_f \rho_f$. Here, $f = r$ with $w_r = 1/3$ corresponds to fast reheating, while $f = \phi$ with $w_\phi = 0$ describes slow reheating.

To compute the density contrast, we employ relativistic perturbation theory. All quantities are decomposed into a homogeneous background and a perturbation, e.g., $
\rho(t,\mathbf{x}) = \bar{\rho}(t) + \delta \rho(t,\mathbf{x})$. The averaged energy density $\bar\rho_f$ can be resolved by the continuity equation $\dot{\bar{\rho}}_f + 3 H (1+w_f)\bar{\rho}_f=-\dot{\bar{\rho}}_V$ together with the first Friedmann equation $H^2=(\bar{\rho}_V+\bar{\rho}_f)/(3m_{\rm Pl}^2)$ with $m_{\rm Pl}\approx2.4\times 10^{18}$ GeV being the reduced Planck scale. For a supercooled FOPT, the initial condition is $\bar\rho_f=0$ and $\bar\rho_V=\Delta V$.

Given the perturbations, the energy density contrast is
\be
\delta =\frac{\delta\rho}{\bar\rho}\equiv\frac{\delta\rho_V+\delta\rho_f}{\bar\rho_V+\bar\rho_f}\,.
\ee
The background spacetime is taken to be the flat Friedmann-Lema\^{\i}tre-Robertson-Walker metric. Restricting to the dominant scalar perturbations, the line element is
\begin{multline}
\d s^2=(1+2A)\,\d t^2-2a(\partial_iB)\,\d t\,\d x^i\\
-a^2\left[\delta_{ij}(1+2C)+2\partial_{\langle i}\partial_{j\rangle}E\right]\d x^i\d x^j\,,
\end{multline}
where $\partial_{\langle i}\partial_{j\rangle}E\equiv(\partial_i\partial_j-\delta_{ij}\nabla^2/3)E$.

{\it Gauge-invariant equations}---Both the density contrast and metric perturbations defined above depend on gauge. To isolate physics from gauge artifacts, we introduce the following gauge-invariant variables: the Bardeen potential $\Phi=-C+\nabla^2E/3-aH(B-a\dot E)$~\cite{Bardeen:1980kt}, the comoving density contrast $\Delta=[\delta\rho+a\dot{\bar\rho}(v+B)]/\bar\rho$, and the velocity variable $\mV=v+ a \dot{E}$. Here $v$ is the velocity potential that generates the velocity perturbation via $\v=\nabla v$. Note $\Delta\equiv \delta^{(C)}$, as the comoving gauge is defined by $v+B=0$ and $E=0$.

Following relativistic perturbation theory, one obtains the gauge-invariant perturbed continuity and Euler equations
\bea\label{continuity}
\dot\Delta_k-3H\frac{\bar P}{\bar\rho}\Delta_k&=&\left(1+\frac{\bar P}{\bar\rho}\right)\frac{k^2\mV_k}{a}\,,\\
\dot\mV_k+H\mV_k&=&-\frac{1}{a(\bar\rho+\bar P)}\left(\delta P_{\rm nad}+\frac{\dot{\bar P}}{\dot{\bar\rho}}\bar\rho\Delta_k\right)-\frac{\Phi_k}{a}\,,\nn
\eea
and the Poisson equation $k^2\Phi_k=-a^2\bar\rho\Delta_k/(2m_{\rm Pl}^2)$. Here the subscript $k$ labels the Fourier mode corresponding to a comoving volume with radius $k^{-1}$, $\bar P=\bar P_V+\bar P_f$ is the total background pressure, and $\delta P_{\rm nad}=\delta P-(\dot{\bar P}/\dot{\bar\rho})\delta\rho$ is the gauge-invariant non-adiabatic pressure perturbation. \Eq{continuity} follows from standard formulas in relativistic perturbation theory (see, e.g., Refs.~\cite{Kodama:1984ziu,baumann2022cosmology,mukhanov2005physical}); a concise derivation is given in the Supplemental Material.

{\it Evaluating $\Delta_k$ by bubble simulations}---Since \Eq{continuity} is gauge-invariant, all variables therein may be evaluated in any gauge. For practical convenience, here we evaluate them in the flat gauge $C=E=0$. Then $\Phi_k=-aHB_k$, and the Poisson equation gives $B_k=3aH\Delta_k/(2k^2)$. Using $\Delta_k$ to express $\delta\rho$, and applying the two-fluid assumption, we get
\begin{multline}
\delta P_{\rm nad}=(1+w_f)\dot{\bar\rho}_V\left(\frac{\bar\rho}{\dot{\bar\rho}}-\frac{3}{2}\frac{a^2H}{k^2}\right)\Delta_k\\
-(1+w_f)\left(a\mV_k\dot{\bar\rho}_V+\delta\rho_V\right)\,.
\end{multline}
Substituting it into the Euler equation, and letting $\mQ_k\equiv (\bar\rho+\bar P)\mV_k$ be the gauge-invariant momentum density, we obtain
\begin{multline}\label{key1}
\dot\mQ_k+(4+3w_f)H\mQ_k=\frac{1+w_f}{a}\delta\rho_V\\
-\left[\frac{w_f}{a}\bar\rho+\frac{aH}{2k^2}\left(\dot{\bar\rho}-3(1+w_f)\dot{\bar\rho}_V\right)\right]\Delta_k\,.
\end{multline}
Meanwhile, the continuity equation reduces to
\be\label{key2}
\dot\Delta_k-3H\frac{\bar P}{\bar\rho}\Delta_k=\frac{k^2\mQ_k}{a\bar\rho}\,.
\ee

\begin{figure}
\centering
\includegraphics[scale=0.45]{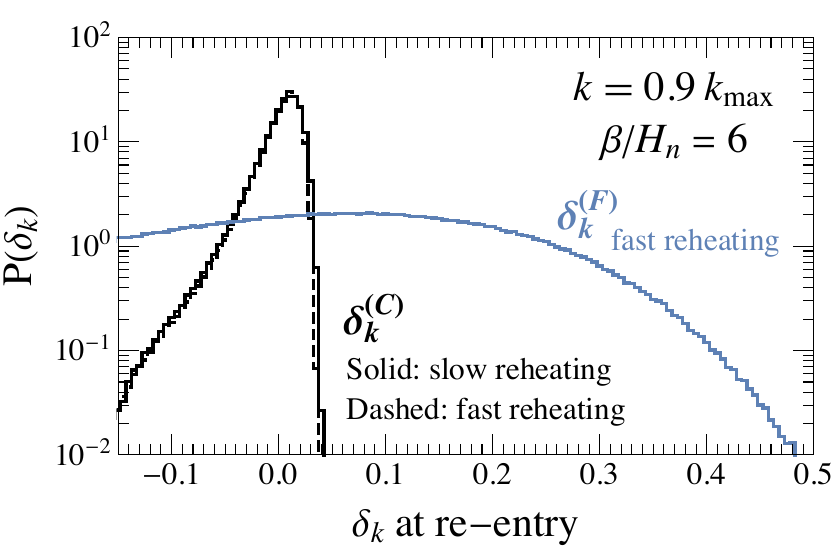}
\caption{Density contrast PDFs in the comoving volume $4\pi k^{-3}/3$ at horizon re-entry, with $k=0.9\,k_{\rm max}$ and $\beta/H_n=6$. The comoving‑ and flat-gauge $\delta_k$'s are shown as black and blue lines, respectively.}\label{fig:deltak}
\end{figure}

Equations~(\ref{key1}) and~(\ref{key2}) form a closed system for the gauge-invariant variables $\Delta_k$ and $\mQ_k$, although the intermediate expression for $\delta P_{\rm nad}$ is evaluated in the flat gauge. We simulate the bubble nucleation history and extract $\delta\rho_V$ by modifying the {\tt deltaPT} code~\cite{Lewicki:2024ghw}: in a spherical comoving volume $4\pi k^{-3}/3$, the first $j_c$ bubbles are generated individually to capture fluctuations in their nucleation times and positions, while the remaining bubbles are treated as an averaged background. We evolve $\Delta_k\equiv\delta_k^{(C)}$ up to the horizon re-entry moment $t_k$. Repeating this for $N_{\rm sim}$ realizations, the PDF of $\delta_k^{(C)}$ at re-entry is obtained. As the PDF shifts to larger values for larger $k$, we adopt $k=0.9\,k_{\rm max}$ as a benchmark.

Figure~\ref{fig:deltak} shows the PDF of $\delta_k^{(C)}$ for $\beta/H_n=6$, $j_c=50$, and $N_{\rm sim}=10^6$, with slow reheating (black solid line) and fast reheating (black dashed line). The PDF of $\delta_k^{(F)}$ from the original {\tt deltaPT} code for fast reheating is shown in a blue line for comparison; while the curve of $\delta_k^{(F)}$ for slow reheating is omitted because our analysis focuses on $\delta_k^{(C)}$ entering the PBH calculation. The density contrast is insensitive to reheating efficiency but highly sensitive to gauge choice: the comoving-gauge values are significantly smaller than the flat-gauge ones, consistent with the literature~\cite{Franciolini:2025ztf,Wang:2026zvz}. Unlike Ref.~\cite{Wang:2026zvz}, which infers $\delta_k^{(C)}$ using the superhorizon relations, our gauge-invariant formulation remains valid without invoking this approximation. We thus provide an independent confirmation of the gauge-dependence issue, showing that the suppression is a general physical effect rather than an artifact of the approximation scheme.

Hereafter, we drop the superscript (C) and let $\delta_k$ denote the comoving-gauge value. The PDF is well fitted as $P(\delta_k)\propto\exp\{\epsilon_k(\delta_k-\mu_k)/2-2\,[1-e^{\epsilon_k(\delta_k-\mu_k)/2}]^2/(\epsilon_k^2\sigma_k^2)\}$~\cite{Lewicki:2024ghw}. 

{\it Reheating after the FOPT}---Most previous works implicitly assume fast reheating, i.e., $\Gamma_\phi/H_*>1$, so that after the FOPT the Universe is instantaneously reheated to a temperature $T_{\rm reh}=T_V$ and enters a radiation-dominated era. Here $T_V$ is given by $\Delta V=\pi^2g_\rho T_V^4/30$, with $g_\rho$ the number of degrees of freedom (dof) for energy. In such a case, radiation pressure resists gravitational collapse; PBH formation therefore demands a large density contrast. Consequently, the relevant quantity is the $\delta_k$ at horizon re-entry, and the smallness of $\delta_k$ we obtain thus rules out PBH formation in this standard scenario.

However, the situation changes qualitatively when reheating is slow, i.e., $\Gamma_\phi < H_*$. After the FOPT, the vacuum energy is transferred into coherent oscillations of the scalar field, which subsequently decay into radiation. The energy densities evolve according to~\cite{Hambye:2018qjv}
\be\label{rhe}
\dot{\bar\rho}_\phi+3H\bar\rho_\phi=-\Gamma_\phi\bar\rho_\phi,\quad \dot{\bar\rho}_r+4H\bar{\rho}_r=\Gamma_\phi\bar\rho_\phi\,,
\ee
with $H^2=(\bar{\rho}_\phi+\bar{\rho}_r)/(3m_{\rm Pl}^2)$ and initial conditions $\bar\rho_\phi\approx \Delta V$, $\bar\rho_r\approx0$. Reheating ends at $t_{\rm reh}$ when $\bar\rho_\phi(t_{\rm reh})=\bar\rho_r(t_{\rm reh})$, and the corresponding temperature is denoted $T_{\rm reh}$. Numerical integration of the above equations gives $T_{\rm reh}\approx0.64\,(\Gamma_\phi/H_*)^{1/2}T_V$, consistent with the analytic approximation~\cite{Hambye:2018qjv}. Between $t_k$ and $t_{\rm reh}$, the Universe is in an EMD era, during which the absence of pressure allows a small $\delta_k$ to grow efficiently and form PBHs.

{\it PBH formation in the EMD era}---In a matter-dominated Universe, the threshold for gravitational collapse is significantly reduced, and the collapse proceeds more gradually. At the same time, the density contrast, small initial deviations from spherical symmetry, and any angular momentum can grow throughout the collapse process. Hence, the collapse in an EMD era is sensitive to the specific shape of the perturbation in the region under consideration, described by a symmetric three-by-three tensor $\mD$. The evolution typically leads to the so-called ``pancake collapse''~\cite{Zeldovich:1969sb,Khlopov:1980mg}.  Therefore, predicting the PBH abundance requires not only the distribution of the density contrast $\delta_k=-{\rm tr}\, \mD$, but also the joint statistics of the independent components of the deformation tensor and the initial angular momentum.

PBH formation in an EMD era was systematically studied in Refs.~\cite{Harada:2016mhb,Harada:2017fjm} under the assumption that all deformation parameters arise from a single Gaussian random scalar field. In our case, however, the simulated $P(\delta_k)$ is manifestly non-Gaussian, so their framework must be extended. We make two assumptions: first, the joint distribution of the symmetric deformation factorizes as
\be\label{factorization}
P_{\mD}(\mD_{ij})=P(\delta_k)P_{\tilde{\mD}}(\tilde{\mD}_{ij})\,,
\ee
where $\tilde{\mD}_{ij}\equiv \mD_{ij}+\delta_k\delta_{ij}/3$ is the traceless part of the deformation tensor; second, $P_{\tilde{\mD}}(\tilde \mD_{ij})$ is isotropic and Gaussian. The PBH formation probability is then obtained as (see the Supplemental Material for details)
\begin{multline}\label{eq:betak-2}
\beta_{k}= f_q(q_c)\int_{-\infty}^\infty\d\delta_k\int_{-\infty}^\infty \d y\int_{2|y|}^\infty\d z\,\theta\left(1-h(\delta_k,y,z)\right)\\
\times P(\delta_k)\frac{9\sqrt{2}}{\sqrt{\pi} \sigma_3^5}(z^3-4zy^2)e^{-\frac{1}{\sigma_3^2} \left(2y^2+\frac{3}{2}z^2\right)}\,.
\end{multline}
The step function enforces the hoop‑conjecture condition $h\leqslant1$, where $h(\delta_k,y,z)=36z\,\mE(\xi)/[\pi(\delta_k+2y+3z)^2]$ with $\xi=\sqrt{1-(z+2y)^2/4z^2}$ and $\mE$ being the complete elliptic integral of the second kind. The parameter $\sigma_3$ controls the width of the distribution of the traceless part and is in principle independent of $P(\delta_k)$; to recover the Gaussian limit~\cite{Harada:2016mhb,Harada:2017fjm}, we set $\sigma_3 = \tilde\sigma_k/\sqrt{5}$ with $\tilde{\sigma}^2_{k}=\overline{\delta_k^2}$, which implicitly correlates the trace and the traceless parts. As in Ref.~\cite{Harada:2017fjm}, $\mI\sim\mO(1)$ and $f_q(q_c)\approx1$ is the fraction of configurations with quadrupole asphericity $q<q_c \simeq 2.4\,(\mI\tilde\sigma_k)^{1/3}$.

The factorization ansatz \Eq{factorization} together with a Gaussian traceless part is a strong but deliberate assumption. Current bubble simulations can only determine the density PDF $P(\delta_k)$, while the full $\mD_{ij}$ remains out of reach, making some closure assumption unavoidable. Our ansatz is the minimal extension of the standard framework: it modifies only the trace part, whose non-Gaussian statistics are directly obtained from simulations, while preserving the conventional Gaussian description of the traceless sector. In the Gaussian limit, it exactly reduces to the standard Doroshkevich distribution used in previous studies. This allows us to provide the first calculable estimate of the PBH formation probability $\beta_k$ sourced by the non-Gaussian density perturbations generated in a slow FOPT. A more complete treatment will require future simulations to reconstruct the full $\mD_{ij}$, thereby testing the validity of the factorization ansatz and determining possible correlations between the trace and traceless components.

During the EMD era, the overdensity first grows, then starts collapsing at $t_{\rm coll}$, given by $a(t_{\rm coll})\approx a_k\cdot (0.4\,\mI\tilde\sigma_k)^{-2/3}$~\cite{Harada:2017fjm}. PBH formation thus requires $t_{\rm reh}>t_{\rm coll}$ or equivalently $a_{\rm reh}\equiv a(t_{\rm reh})>a(t_{\rm coll})$. To translate this into a bound on $T_{\rm reh}$, we relate $a_{\rm reh}$ to the temperature. Assuming entropy conservation after $t_{\rm reh}$ and normalizing the scale factor today to $a_0=1$ with temperature $T_0\approx 2.35\times10^{-13}$ GeV, one has $a_{\rm reh}^3g_s(T_{\rm reh})T_{\rm reh}^3 = g_s(T_0)T_0^3$, where $g_s$ denotes the number dof for entropy. The scale factor $a_k$ can be expressed through the matter-era scaling $(a_k/a_{\rm reh})^3=(H_{\rm reh}/H_k)^2$ using $H_k = k/a_k$ and $H_{\rm reh}^2\equiv H^2(t_{\rm reh})=2\pi^2g_\rho(T_{\rm reh}) T^4_{\rm reh}/(90 m_{\rm Pl}^2)$ from matter-radiation equality. Combining these relations yields the PBH formation condition as an upper limit on the reheating temperature: $T_{\rm reh}\leqslant T_{\rm reh}^{\rm max}$, where
\be\label{PBH_criterion}
T_{\rm reh}^{\rm max}=\left(0.4\,\mathcal{I}\tilde\sigma_k\frac{g_s(T_{\rm reh})}{g_s(T_0)}\right)^{1/3}\!\!\sqrt{\frac{45}{\pi^2 g_\rho(T_{\rm reh})}}\frac{k\, m_{\rm Pl}}{T_0}\,.
\ee
A sufficiently small $\Gamma_\phi/H_*$ prolongs the EMD era, lowering $T_{\rm reh}$, and thereby allowing PBH formation.

{\it The PBH profile}---PBHs formed in an EMD era may undergo significant accretion, greatly enhancing their mass~\cite{DeLuca:2021pls}. Given the uncertainties in the accretion process, we bracket the possibilities by two extreme cases: no accretion, where the PBH mass is set at the moment of horizon re-entry~\cite{Ballesteros:2019hus}, and maximal accretion (adopted e.g.~in Ref.~\citep{Ai:2024cka}), where the PBH can grow up to the Hubble mass at the end of the EMD~\cite{deJong:2021bbo}. The final PBH mass therefore satisfies $M_{\rm min}\equiv4\pi\gamma\, m^2_{\rm Pl}/H_k <M_{\rm pbh}<M_{\rm max}\equiv 4\pi\gamma\,m^2_{\rm Pl}/H_{\rm reh}$. Here, the parameter $\gamma\sim\mO(1)$ quantifies the collapse efficiency, and for simplicity we set it to 1 below. Using the scaling relations derived above, we have
\be\label{Mpbh}\begin{split}
M_{\rm min}=&~\frac{4\pi^3}{45}\left(\frac{T_0}{k}\right)^3 \frac{g_s(T_0)}{g_s(T_{\rm reh})} g_\rho(T_{\rm reh}) T_{\rm reh}\,,\\
M_{\rm max}=&~\frac{4m_{\rm Pl}^3}{T_{\rm reh}^2}\sqrt{\frac{45}{g_\rho(T_{\rm reh})}}\,.
\end{split}\ee

The quantity $\beta_{k}$ gives the fraction of Hubble patches at horizon re-entry $t_k$ that eventually collapse into PBHs~\cite{Harada:2017fjm}. The resulting PBH number density at the end of reheating is
\be
n_{\rm pbh}(t_{\rm reh})=\frac{\beta_{k}}{4\pi H_k^{-3}/3}\left(\frac{a_k}{a_{\rm reh}}\right)^3=\frac{3\beta_{k}}{4\pi}H_{\rm reh}^2H_k\,.
\ee
To connect with the current PBH abundance, we define the yield $Y_{\rm pbh} \equiv n_{\rm pbh}/s=n_{\rm pbh}(t_{\rm reh})/[2\pi^2g_s(T_{\rm reh})T_{\rm reh}^3/45]$, which is unchanged after reheating.

The present‑day PBH fraction of dark matter is
\be
f_{\rm pbh}=\frac{\Omega_{\rm pbh}}{\Omega_{\rm dm}}=\frac{1}{\Omega_{\rm dm}h^2}\frac{M_{\rm pbh}Y_{\rm pbh}s_0}{3m_{\rm Pl}^2}\left(\frac{h}{H_0}\right)^2\,,
\ee
where $s_0\approx2891.2~{\rm cm}^{-3}$ is the entropy density today, $\Omega_{\rm dm}h^2=0.12$, and $H_0/h=100~{\rm km}/({\rm s}\cdot{\rm Mpc})$~\cite{ParticleDataGroup:2024cfk}. Inserting the minimal and maximal PBH masses gives the corresponding fraction bounds:
\be\label{fpbh}\begin{split}
f_{\rm min}=&~\frac{\beta_{k}}{2\,\Omega_{\rm dm}h^2}\left(\frac{s_0T_{\rm reh}}{m_{\rm Pl}^2}\right)\frac{g_\rho(T_{\rm reh})}{g_s(T_{\rm reh})}\left(\frac{h}{H_0}\right)^2\,,\\
f_{\rm max}=&~\frac{\beta_{k}}{2\,\Omega_{\rm dm}h^2}\left(\frac{m_{\rm Pl}s_0}{T_{\rm reh}^2}\right)
\frac{g_\rho(T_{\rm reh})}{g_s(T_0)}\left(\frac{h}{H_0}\right)^2\\
&~\times\left(\frac{45}{\pi^2g_\rho(T_{\rm reh})}\right)^{3/2}\left(\frac{k}{T_0}\right)^3\,.
\end{split}\ee

\begin{figure}
\centering
\includegraphics[scale=0.45]{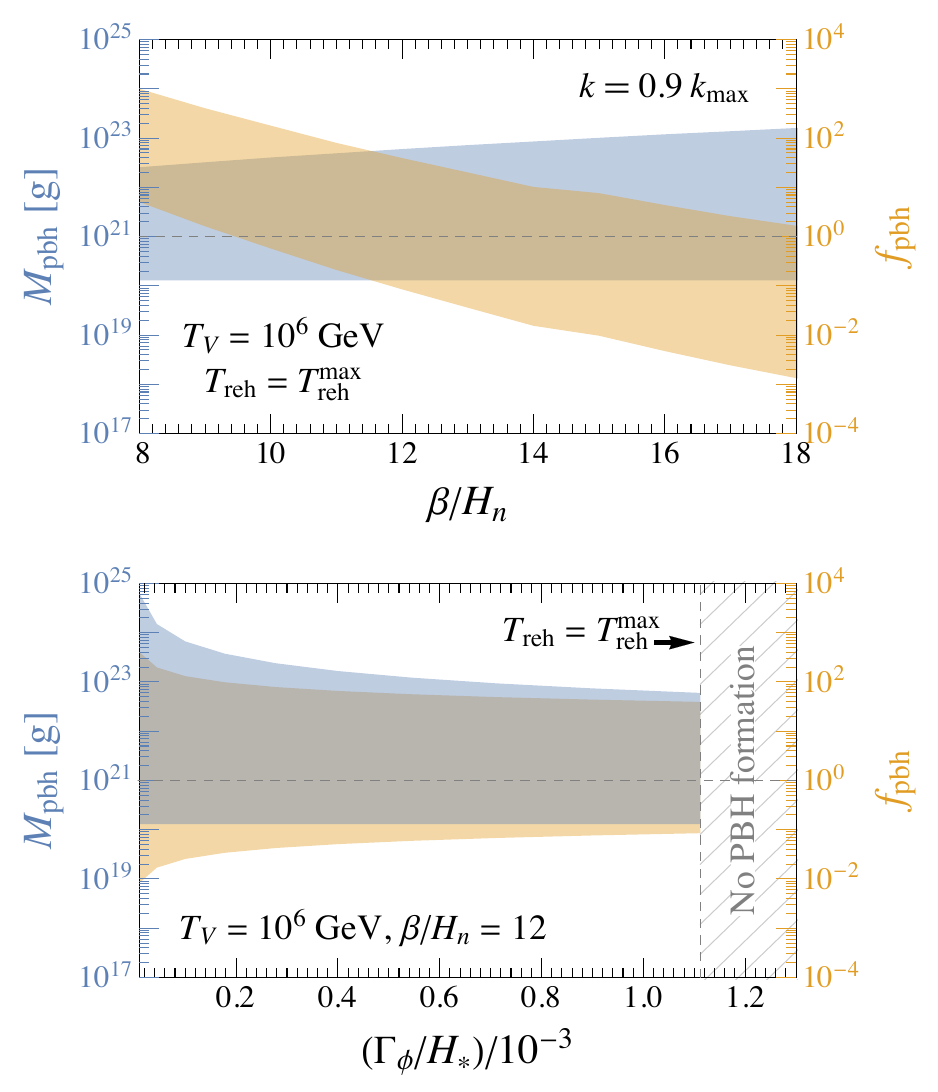}
\caption{PBH profile for $k=0.9\,k_{\rm max}$ and $T_V=10^6$ GeV. Upper: fixing $T_{\rm reh}=T_{\rm reh}^{\rm max}$, and varying $\beta/H_n$. Lower: fixing $\beta/H_n=12$, and varying $\Gamma_\phi/H_*$.}\label{fig:pbh}
\end{figure}

Figure~\ref{fig:pbh} shows $M_{\rm pbh}$ (blue band) and $f_{\rm pbh}$ (orange band) for $k=0.9\,k_{\rm max}$ and $T_V=10^6$ GeV. The upper panel fixes $T_{\rm reh}=T_{\rm reh}^{\rm max}$ while varies $\beta/H_n$ from 8 to 18. $M_{\rm pbh}$ lies within $10^{20}$ g -- $10^{22}$ g, an ``asteroid-mass'' window that can explain all of the dark matter while satisfying current constraints~\cite{Carr:2020gox,Carr:2020xqk,Carr:2026hot}, and that could be probed in the future by femtolensing, gamma-ray burst lensing parallax, etc~\cite{Katz:2018zrn,Bai:2018bej,Jung:2019fcs,Gawade:2023gmt,Fedderke:2024wpy,Kaplan:2024dsn}. $f_{\rm pbh}$ changes by roughly five orders of magnitude, showing a strong dependence on $\beta/H_n$. The lower panel fixes $\beta/H_n=12$ and scans over $\Gamma_\phi/H_*$, thereby varying $T_{\rm reh}$. PBH formation occurs only for $T_{\rm reh}\leqslant T_{\rm reh}^{\rm max}$, and $f_{\rm pbh}$ displays a milder dependence on $\Gamma_\phi/H_*$ than on $\beta/H_n$. Here we set $N_{\rm sim}=10^6$ with $j_c=50$ (80) for $\beta/H_n<15$ ($\geqslant15$); using $N_{\rm sim}=10^5$ already yields nearly identical results.

We also find that $\beta_k$ is maximized at $k/k_{\rm max}\approx 0.7-0.9$, leading to a quasi-monochromatic mass spectrum for fixed model parameters. Accretion primarily shifts the peak mass without significantly broadening the distribution. We then use $k=0.9\,k_{\rm max}$ as a benchmark.

We briefly comment on PBH spin. During an EMD era, efficient angular momentum accumulation may yield a sizable Kerr parameter $\chi \equiv 8\pi J_{\rm pbh} (m_{\rm Pl}/M_{\rm pbh})^2$, where $J_{\rm pbh}$ is the PBH angular momentum. Ref.~\cite{Harada:2017fjm} finds a statistical anti-correlation $\chi\propto\delta_k^{-3/2}$, implying that PBHs are typically born with near-extremal spins ($\chi\sim1$), since their formation is dominated by perturbations near the collapse threshold. However, Ref.~\cite{Ye:2025wif} points out that this anti-correlation holds for the ensemble of all perturbations, not those that collapse into PBHs. Restricting to perturbations that actually form PBHs gives a positive correlation between $\chi$ and $\delta_k$, leading to typically low PBH spins. While the analysis of Ref.~\cite{Ye:2025wif} assumes Gaussian perturbations, we expect that our non-Gaussian density contrast would also result in low spins.

{\it Analytic estimates}---We now provide convenient analytic expressions for the PBH profile. Using the relation $a_*= a_{\rm reh}\cdot (H_{\rm reh}/H_*)^{2/3}$ from the EMD era, and the approximation $H_*^2\approx \Delta V/(3 m^2_{\rm Pl})$ for a supercooled FOPT, $k_{\rm max}$ can be written in terms of $T_{\rm reh}$ and $\Delta V$.  For a simple order-of-magnitude estimate, we take $g_\rho(T_{\rm reh})\approx g_s(T_{\rm reh})\approx110$. Numerical results show $\tilde\sigma_k\approx0.006\times10^{0.07\,(12-\beta/H_n)}$ and $\beta_{k}\approx10^{-15}\times 10^{0.3\,(12-\beta/H_n)}$ within the range $8\lesssim\beta/H_n\lesssim18$ for $k=0.9\,k_{\rm max}$. With these inputs, the PBH formation condition \Eq{PBH_criterion} simplifies to a bound on reheating efficiency, $\Gamma_\phi/H_*\lesssim0.004\times10^{0.07\,(12-\beta/H_n)}$.

The corresponding PBH mass and dark matter fraction can also be estimated analytically.
For the case without accretion,
\bea
M_{\rm min}&\sim&10^{20}~{\rm g}\times\left(\frac{10^6~{\rm GeV}}{T_V}\right)^2\,,\\
f_{\rm min}&\sim&2\times 10^{0.3\,(12-\beta/H_n)}\left(\frac{T_V}{10^6~{\rm GeV}}\right)\left(\frac{\Gamma_\phi}{H_*}\right)^{1/2}\,,\nn
\eea
while for maximal accretion,
\bea
M_{\rm max}&\sim&2\times10^{20}~{\rm g}\times\left(\frac{10^6~{\rm GeV}}{T_V}\right)^2\left(\frac{H_*}{\Gamma_\phi}\right)\,.\\
f_{\rm max}&\sim&4\times10^{0.3\,(12-\beta/H_n)}\left(\frac{T_V}{10^6~{\rm GeV}}\right)\left(\frac{H_*}{\Gamma_\phi}\right)^{1/2}.\nn
\eea
The above expressions capture the numerical trends in Fig.~\ref{fig:pbh}. More accurate fit formulas are in the Supplemental Material.

{\it Model-building guide}---Our mechanism requires two essential ingredients: a supercooled and slow FOPT, and sufficiently slow reheating after the transition. The first condition is naturally realized in classically conformal (CC) theories, originally motivated by the hierarchy problem~\cite{Bardeen:1995kv}. In these models, the FOPT scalar is typically embedded in a complex field $\Phi=(\phi+i\eta)/\sqrt{2}$, whose symmetry breaking is driven radiatively through the Coleman-Weinberg mechanism~\cite{Coleman:1973jx}. The resulting one-loop potential,
\begin{equation}
V_1(\phi)=\frac{\mB}{4}\phi^4\left(\log\frac{\phi}{w}-\frac14\right),
\label{V1}
\end{equation}
naturally gives rise to a supercooled and slow FOPT in the early Universe~\cite{Witten:1980ez}.

The second condition requires $\phi$ to couple weakly to the Standard Model (SM), such that its decay rate satisfies $\Gamma_\phi\lesssim H_*$. This is naturally realized in a broad class of models where high-scale CC symmetry breaking triggers electroweak symmetry breaking, such as the minimal $B-L$ model~\cite{Iso:2009ss}. Here $\Phi$ carries $B-L$ charge $+2$ with $\mB=(6/\pi^2)(g_{B-L}^4-\sum_i y_i^4/96)$, where $y_i$ denotes the Yukawa coupling to the right-handed neutrinos. The Higgs-$\phi$ mixing angle, $\theta\approx(v_{\rm ew}/w)m_h^2/(m_h^2-m_\phi^2)$, is naturally suppressed by the hierarchy $w\gg v_{\rm ew}$, leading to $\Gamma_\phi\propto\theta^2\ll H_*$~\cite{Hambye:2018qjv,Athron:2025pog}. As another example, consider a dark $U(1)_X$ gauge theory, where $\Phi$ carries charge $+1$ and $\mB=3g_X^4/(8\pi^2)$. The dark sector couples to the SM only through kinetic mixing, $\mL_{\rm int}=-(\sin\epsilon/2)X_{\mu\nu}B^{\mu\nu}$~\cite{Babu:1997st,Chun:2010ve}, yielding the parametrically suppressed decay width $\Gamma_\phi\propto\epsilon^4\ll H_*$. A benchmark realizing this scenario is presented in the Supplemental Material.

The above two ingredients are controlled by different sectors of the theory: the supercooled and slow FOPT is determined by the scalar potential, whereas the EMD duration is governed by the decay rate. Moreover, while the PBH abundance depends exponentially on the FOPT duration, its dependence on the reheating duration is only power-law, making the mechanism much less sensitive to the latter.

{\it Conclusion}---In this Letter, we revisited PBH formation from slow FOPTs. Using a new gauge-invariant treatment independent of the superhorizon approximation, we confirmed the recently identified gauge-dependence issue: the comoving density contrast is substantially smaller than the flat-gauge value. The central insight of this work, however, is that this does {\it not} preclude PBH formation. We have shown that if the post-FOPT reheating is slow, the Universe undergoes an EMD era, during which even the relatively small overdensities can grow efficiently and collapse into PBHs. 

We derived the PBH formation conditions in terms of the reheating duration, and computed the resulting PBH profile. Moreover, the slow and supercooled FOPT that produces the PBHs also generates a strong stochastic GW background, and the subsequent EMD may leave a further spectral imprint on it~\cite{Barenboim:2016mjm,DEramo:2019tit,Ellis:2020nnr,Athron:2025pog}. In addition, the non-spherical collapse in the EMD era itself could be an independent GW source~\cite{Escriva:2026blk}. On the other hand, since our comoving-gauge density contrast agrees with the recent analyses~\cite{Franciolini:2025ztf,Wang:2026zvz}, we expect the associated scalar-induced GW signal to be similarly suppressed. GWs together with the PBH observables serve as a defining signature of this scenario. Our mechanism is general and can be naturally realized in a wide class of new physics models, particularly those with feeble couplings between the dark sector and the SM.

Finally, we comment on the onset of the EMD era. Recently, a lattice simulation work~\cite{Mansour:2026sdx} pointed out that bubble collisions may leave behind persistent relativistic gradient energy, delaying the onset of coherent scalar oscillations. This suggests that an immediate transition to EMD after a supercooled FOPT depends on the nonequilibrium post-collision dynamics rather than being automatic. Our analysis applies to the prompt thermalization scenarios in which the gradient energy is efficiently converted into matter-like coherent oscillations on a sufficiently short timescale. Whether this occurs in concrete models remains an open question that calls for dedicated lattice simulations.\\

{\it Acknowledgements}---We would like to thank Yann Gouttenoire, Zhaofeng Kang, Kazunori Kohri, Jing Liu, Henda Mansour, Chi Tian, and Chen Yuan for helpful discussions and comments. KPX thanks the hospitality of the Tsung-Dao Lee Institute, where part of this work was carried out. WA is supported by startup funds from the Tsung-Dao Lee Institute and Shanghai Jiao Tong University. KPX is supported by the National Natural Science Foundation of China under Grant No.~12305108. The authors are grateful to the Mainz Institute for Theoretical Physics (MITP) of the Cluster of Excellence PRISMA$^+$ (Project ID 390831469), for its hospitality and its partial support during the completion of this work.

\bibliographystyle{utphys}
\bibliography{references}

\onecolumngrid

\pagebreak

\titlepage
\setcounter{page}{1}
\makeatletter

\clearpage
\setcounter{footnote}{0}
\renewcommand{\thefootnote}{\alph{footnote}}

\setcounter{equation}{0}
\renewcommand{\theequation}{S\arabic{equation}}

\begin{center}
{\Large\textbf{Supplemental Material}}
\end{center}

\begin{center}
{\large\textit{Reviving primordial black hole formation in slow first-order phase transitions}}
\end{center}

\begin{center}
{\large Wen-Yuan Ai\textsuperscript{1,2,}\footnote{wenyuanai@sjtu.edu.cn} and Ke-Pan Xie\textsuperscript{3,}\footnote{kpxie@buaa.edu.cn}}\\[6pt]
\textsuperscript{1}\textit{State Key Laboratory of Dark Matter Physics,
  Tsung-Dao Lee Institute and School of Physics and Astronomy,
  Shanghai Jiao Tong University, Shanghai 201210, China}\\[2pt]
\textsuperscript{2}\textit{Key Laboratory for Particle Astrophysics and Cosmology (MOE),
  and Shanghai Key Laboratory for Particle Physics and Cosmology,
  Shanghai Jiao Tong University, Shanghai 201210, China}\\[2pt]
\textsuperscript{3}\textit{School of Physics, Beihang University, Beijing 100191, China}
\end{center}

\section{Derivation of the gauge-invariant equations}

Here we re‑express the standard Newton‑gauge equations in terms of gauge‑invariant variables; the resulting equations then hold in any gauge. In the Newton gauge ($B=E=0$), the gauge‑invariant variables reduce to
\be
\Phi=-C+\frac13\nabla^2E,\quad \Delta=\frac{\delta\rho+a\dot{\bar\rho}v}{\bar\rho},\quad \mV=v\,.\quad\text{(Newton gauge)}
\ee
The scalar part of the perturbed Euler equation in this gauge~\cite{baumann2022cosmology,mukhanov2005physical},
\be
\dot v+H\left(1-3\frac{\dot{\bar P}}{\dot{\bar\rho}}\right)v=-\frac{\delta P}{a(\bar\rho+\bar P)}-\frac{\Phi}{a},\quad\text{(Newton gauge)}
\ee
can be rewritten by substituting the gauge‑invariant combinations above. This yields
\be\label{GI_Euler}
\dot\mV+H\mV=-\frac{1}{a(\bar\rho+\bar P)}\left(\delta P_{\rm nad}+\frac{\dot{\bar P}}{\dot{\bar\rho}}\bar\rho\Delta\right)-\frac{\Phi}{a}\,.
\ee
For the perturbed continuity equation~\cite{baumann2022cosmology,mukhanov2005physical},
\be\label{perturbed_continuity}
\delta\dot\rho+3H(\delta\rho+\delta P)+\frac{\bar\rho+\bar P}{a}\nabla^2v-3(\bar\rho+\bar P)\dot\Phi=0\,,\quad\text{(Newton gauge)}
\ee
we rewrite the expression using the background continuity equation in the main text and the Poisson equation $\nabla^2\Phi=a^2\bar\rho\Delta/(2m_{\rm Pl}^2)$, converting all terms to gauge‑invariant variables. The result is
\be\label{GI_continuity}
\dot\Delta-3H\frac{\bar P}{\bar\rho}\Delta=-\left(1+\frac{\bar P}{\bar\rho}\right)\frac{\nabla^2\mV}{a}\,.
\ee
Eqs.~(\ref{GI_continuity}) and (\ref{GI_Euler}) are respectively the first and second lines of \Eq{continuity} in the main text.

\section{Calculation of PBH formation probability $\beta_k$}

\subsection{The Zel'dovich approximation}

Consider a fluid in a matter-dominated universe. In the Zel'dovich approximation~\cite{Zeldovich:1969sb}, the Euler coordinates $\{r_i\}$ are 
\begin{align}
r_i=a(t) q_i + b(t) p_i(q_j)\,,
\end{align}
with $a(t)$ the scale factor, $q_i$ the Lagrangian coordinates, and $p_i$ the deviation vector. $b(t)$ denotes a linearly growing mode in an EMD era, which is related to the density contrast as we will see shortly.  $a(t)$ satisfies $
H^2=\bar{\rho}/(3m_{\rm Pl}^2)$ with $\bar{\rho}\propto a^{-3}$. The deformation tensor is then defined as
\begin{align}
D_{ij}\equiv \frac{\partial r_i}{\partial q_j} = a(t)\delta_{ij} +b(t)\frac{\partial p_i}{\partial q_j}\equiv  a(t)\delta_{ij} +b(t) \mD_{ij}\,,
\end{align}
where we have defined a second deformation tensor $\mD_{ij}\equiv \partial p_i/\partial q_j$. We can diagonalize the tensor $\mD$ by
\begin{align}
\mD=R \begin{pmatrix}
-\lambda_1 & 0 & 0 \\
0 & -\lambda_2 & 0 \\
0 & 0 & -\lambda_3
\end{pmatrix} R^T\,,
\end{align}
where $R\in SO(3)$.

We order the eigenvalues such that $\lambda_1\geqslant \lambda_2\geqslant\lambda_3$. To ensure that the perturbation will collapse at least along one of the three axes, $\lambda_1$ is required to be positive. From the mass conservation, we have
\begin{align}
\bar{\rho} a^3 \d^3 q= \d m =\rho \d^3 r = \rho (\det D)\d^3 q= \rho [(a-\lambda_1 b)(a-\lambda_2 b)(a-\lambda_3 b)]\d^3 q\,.
\end{align}
Therefore
\begin{align}
\rho =\bar{\rho} \frac{a^3}{(a-\lambda_1 b)(a-\lambda_2 b)(a-\lambda_3 b)}\,,\quad\Rightarrow\quad
\delta_k(t)\equiv \frac{\rho -\bar{\rho}}{\bar{\rho}}=(\lambda_1+\lambda_2+\lambda_3)\frac{b(t)}{a(t)}\,.
\end{align}
Since the density perturbation in the linear regime grows as $\delta_k(t)\propto a(t)$, we have $b(t)\propto a(t)^2$. Without loss of generality, one can always set $b(t_i)=a(t_i)$, such that 
\begin{align}
\delta_k(t_i)= \lambda_1+\lambda_2+\lambda_3=-{\rm tr} \mD\,.
\end{align}

A detailed analysis on how a PBH forms in an EMD era under an initial density contrast $\delta_k(t_i)$ is given in Ref.~\cite{Harada:2016mhb}. Considering only the anisotropy effect, the PBH production probability can be written as 
\begin{align}
\beta_k=\int_0^\infty\d \lambda_1 \int_{-\infty}^{\lambda_1}\d\lambda_2\int_{-\infty}^{\lambda_2} \d \lambda_3\, \theta(1-h(\lambda_1,\lambda_2,\lambda_3))\times w(\lambda_1,\lambda_2,\lambda_3)\,.
\end{align}
Above,
\begin{align}
h(\lambda_1,\lambda_2,\lambda_3)=\frac{2}{\pi} \frac{\lambda_1-\lambda_3}{\lambda_1^2} \mE\left(\sqrt{1-\left(\frac{\lambda_1-\lambda_2}{\lambda_1-\lambda_3}\right)^2}\right)\,,
\end{align}
where we recall $\mE(x)$ is the complete elliptic integral of the second kind.
The function $\theta(1-h(\lambda_1,\lambda_2,\lambda_3))$ is due to the black hole formation criterion by the hoop conjecture. The variables $\{x,y,z\}$ used in the main text are related to $\{\lambda_1,\lambda_2,\lambda_3\}$ via 
\begin{align}
\label{eq:transf}
  x=\frac{\lambda_1+\lambda_2+\lambda_3}{3}=\frac{\delta_k}{3}\,,\quad y=\frac{(\lambda_1-\lambda_2)-(\lambda_2-\lambda_3)}{4}\,,\quad z=\frac{\lambda_1-\lambda_3}{2}\,.  
\end{align}
The $w(\lambda_1,\lambda_2,\lambda_3)\d\lambda_1 \d\lambda_2\d\lambda_3$ is Doroshkevich's probability distribution~\cite{doroshkevich1970spatial}, derived by assuming that the probability distribution of the independent components of the deformation tensor $\mD_{ij}$ arises from a Gaussian random scalar field. Since the perturbations generated from an FOPT do not satisfy this assumption, we will have to modify the Doroshkevich probability distribution accordingly.

\subsection{Standard Doroshkevich probability distribution}

Here we derive the standard Doroshkevich distribution, reproducing the expressions given in Refs.~\cite{Harada:2016mhb,Harada:2017fjm}. The deformation tensor $\mD$ is symmetric and hence has six independent components; they are labeled as $\{\mD_{11},\mD_{22},\mD_{33},\mD_{12},\mD_{23},\mD_{13}\}\equiv U$. We are interested in $P_{\mD}(\mD_{ij})\d \mD$ where $P_{\mD}(\mD_{ij})$ denotes the probability distribution for the independent components and $\d \mD$ denotes the volume element in the corresponding six-dimensional space. Specifically, $\d \mD=\prod_{\mD_{ij}\in U} \d \mD_{ij}$.
Writing $\mD=R\Lambda R^T$ with $\Lambda={\rm diag}(-\lambda_1,-\lambda_2,-\lambda_3)$ and $R\in SO(3)$, the differential becomes
\begin{align}
\d \mD_{ij} &= (\d R_{im})\Lambda_{mn} (R^T)_{nj} + R_{im}(\d\Lambda_{mn})(R^T)_{nj} + R_{im}\Lambda_{mn} \d (R^T)_{nj}\notag\\
&= R_{im}(\d\Lambda_{mn}+[\d\Omega, \Lambda]_{mn})(R^T)_{nj}\,,
\end{align}
where $\d\Omega= R^T\d R$. Since the measure is rotationally invariant, the volume element factors as
\begin{align}
    \d \mD=\prod_{{\rm independent\  components}} (\d\Lambda_{mn}+[\d\Omega, \Lambda]_{mn}).
\end{align}
The three diagonal contributions are simply $\d \lambda_1$, $\d \lambda_2$, $\d\lambda_3$. For the off‑diagonal modes $m\neq n$,
\begin{align}
    [\d\Omega,\Lambda]_{mn}=(\lambda_n-\lambda_m)(\d\Omega)_{mn}.
\end{align}
Assembling all components gives
\begin{align}
    \d \mD= (\lambda_1-\lambda_2)(\lambda_1-\lambda_3)(\lambda_2-\lambda_3)\d\lambda_1\d\lambda_2\d\lambda_3\d\mu(R)\,,
\end{align}
where $\mu(R)$ is the Haar measure on $SO(3)$. 

Now, we assume that the deformation tensor is induced by a Gaussian scalar field $S$, $\mD_{ij}=\partial_i\partial_j S$. All derivatives of a Gaussian scalar field are also Gaussian. Therefore,
\begin{align}
    P_{\mD}(\mD_{ij})\propto e^{-\frac{1}{2}\mD_{ij} (C^{-1})_{ijkl} \mD_{kl}},
\end{align}
with the covariance tensor
\begin{align}
    C_{ijkl}=\langle \mD_{ij} \mD_{kl}\rangle=\frac{\sigma_3^2}{3}(\delta_{ij}\delta_{kl}+\delta_{ik}\delta_{jl}+\delta_{il}\delta_{jk})\,.
\end{align}
Inverting $C$, imposing normalization, and converting the measure to the eigenvalue parametrization, we finally obtain
\begin{align}
\label{eq:standard-Do}
    P_{\mD}(\mD_{ij})\d \mD 
    &=\frac{27}{8\sqrt{5}\pi \sigma_3^6} (\lambda_1-\lambda_2)(\lambda_1-\lambda_3)(\lambda_2-\lambda_3)\, e^{-\frac{3}{5\sigma_3^2}\left[(\lambda_1^2+\lambda_2^2+\lambda_3^2)-\frac{1}{2}(\lambda_1\lambda_2+\lambda_1\lambda_3+\lambda_2\lambda_3)\right]}\d\lambda_1\d\lambda_2\d\lambda_3\\
    &\equiv w(\lambda_1,\lambda_2,\lambda_3)\d\lambda_1\d\lambda_2\d\lambda_3\,,\notag
\end{align}
where $w(\lambda_1,\lambda_2,\lambda_3)$ is the Doroshkevich probability distribution.

\subsection{Modified Doroshkevich probability distribution used in our calculation}

As discussed in the main text, the distribution of density contrast $\delta_k=-{\rm tr}\mD$ arising from a FOPT is generally non-Gaussian. We therefore proceed by treating its distribution $P_{\mD}(\delta_k)$ as arbitrary and writing the joint probability of the deformation tensor in a factorized form
\begin{align}  P_{\mD}(\mD_{ij})=P(\delta_k)P_{\tilde{\mD}}(\tilde{\mD}_{ij})\,,
\end{align}
where $\tilde{\mD}_{ij}\equiv \mD_{ij}+\delta_k\delta_{ij}/3$ is the traceless part of $\mD$. Our first key assumption is that the statistics of $\tilde{\mD}_{ij}$ remain Gaussian and isotropic. With this assumption,
\begin{align}
    P_{\mD}(\mD_{ij})=\mathcal{N} P(\delta_k)\,e^{-\frac{3}{4\sigma_3^2} {\rm tr}\tilde{\mD}^2}\,,
\end{align}
where $\mathcal{N}$ is a normalization constant. Expressed in terms of the eigenvalues, the volume element becomes
\begin{align}
    P_{\mD}(\mD_{ij})\d \mD=\mathcal{N} P(\delta_k)\,e^{-\frac{3}{4\sigma_3^2}\left[\left(\lambda_1-\frac{\delta_k}{3}\right)^2+\left(\lambda_2-\frac{\delta_k}{3}\right)^2+\left(\lambda_3-\frac{\delta_k}{3}\right)^2\right]}(\lambda_1-\lambda_2)(\lambda_2-\lambda_3)(\lambda_1-\lambda_3)\d\lambda_1\d\lambda_2\d\lambda_3\,,
\end{align}
where $\d \mu(R)$ has been absorbed into the normalization factor. 

Performing the variable transformation~\eqref{eq:transf} and imposing the normalization condition yields
\begin{align}
\label{eq:modified-Do1}
    P_{\mD}(\mD_{ij})\d \mD=P(\delta_k)\times\frac{9\sqrt{2}}{\sqrt{\pi}\sigma_3^5}(z^3-4 zy^2)\, e^{-\frac{1}{\sigma_3^2}\left[2y^2+\frac{3}{2}z^2\right]}\d\delta_k\d y\d z\,.
\end{align}
In principle, the width parameter $\sigma_3$ entering this expression is independent of $P(\delta_k)$. Present simulations, however, only give access to $P(\delta_k)$; a full evaluation therefore requires relating $\sigma_3$ to a quantity that can be extracted from $P(\delta_k)$. This is equivalent to assuming a correlation between the statistics of the trace and the traceless part of the deformation tensor. We fix $\sigma_3$ by demanding that \Eq{eq:modified-Do1} reduces to the standard Doroshkevich form \Eq{eq:standard-Do} when $P(\delta_k)$ is Gaussian, which leads to $\sigma_3 = \tilde{\sigma}_k/\sqrt{5}$, where $\tilde{\sigma}_k^2 \equiv \overline{\delta_k^{2}}$.

\section{Fits for the $\tilde\sigma_k$ and $\beta_{k}$ parameters}

\begin{figure}
\centering
\includegraphics[scale=0.45]{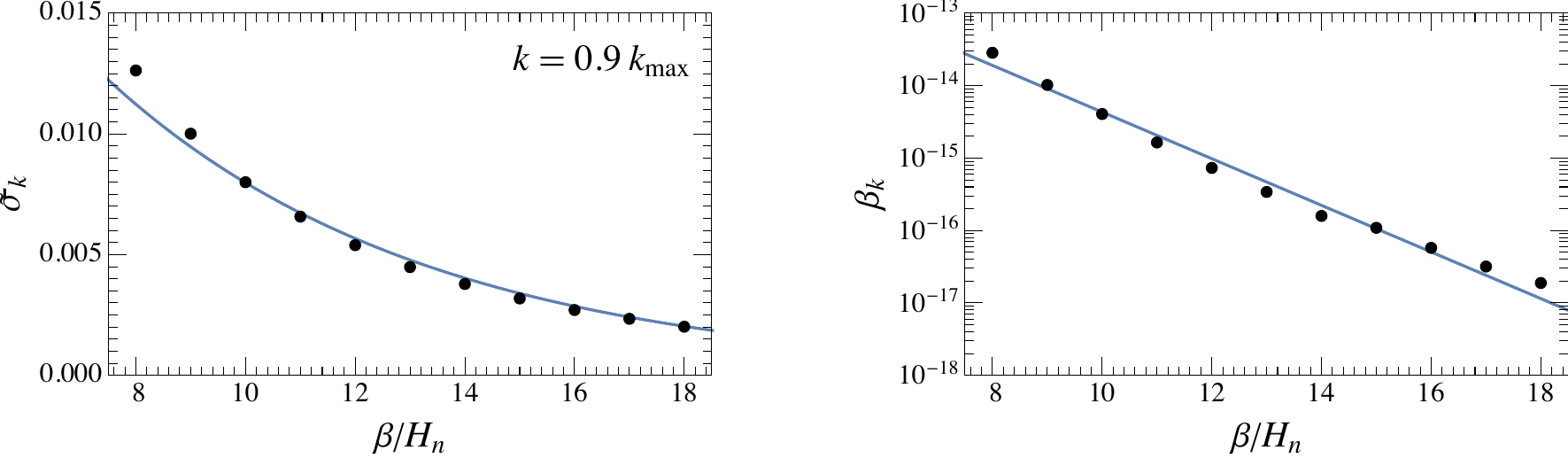}
\caption{$\beta/H_n$-dependence of $\tilde\sigma_k$ (left) and $\beta_{k}$ (right) for $k=0.9\,k_{\rm max}$. Black dots represent the numerical calculation, while blue lines are the fits.}\label{fig:fit}
\end{figure}

Figure~\ref{fig:fit} shows the parameters $\tilde\sigma_k$ and $\beta_{k}$ as functions of $\beta/H_n$ for $k=0.9\,k_{\rm max}$. Here we show the numerical fits,  
\be
\tilde\sigma_k\approx0.00567\times 10^{0.0742\, (12 - \beta/H_n)},\quad \beta_{k}\approx 9.81\times 10^{-16}\times 10^{0.322 \, (12 - \beta/H_n)}\,,
\ee
displayed as blue lines. Together with Eqs.~(\ref{PBH_criterion}), (\ref{Mpbh}), and (\ref{fpbh}) in the main text, these allow a rapid evaluation of the formation condition and profile of the PBH.

\section{Details of the classically conformal $U(1)_X$ model}

The kinetic and mass terms can be diagonalized by a non-unitary transform~\cite{Babu:1997st,Chun:2010ve}
\be
\begin{pmatrix}B_\mu \\ W_\mu^3 \\ X_\mu\end{pmatrix}= U_{\rm kin}\begin{pmatrix}A_\mu \\ Z_\mu \\ \tilde X_\mu\end{pmatrix},\quad U_{\rm kin}=\begin{pmatrix}c_W&-(t_\epsilon s_\xi+s_W c_\xi)&s_Ws_\xi-t_\epsilon c_\xi\\ s_W & c_Wc_\xi & -c_Ws_\xi \\ 0 & s_\xi/c_\epsilon & c_\xi/c_\epsilon\end{pmatrix}\,,
\ee
where $c_W=g/\sqrt{g^2+g'^2}$, $s_W=g'/\sqrt{g^2+g'^2}$, $c_\epsilon=\cos\epsilon$, $t_\epsilon=\tan\epsilon$, $c_\xi=\cos\xi$, $s_\xi=\sin\xi$, and $\xi$ is determined by
\be
\tan2\xi=\frac{-2s_\epsilon c_\epsilon g'\sqrt{g^2+g'^2}v_{\rm ew}^2}{4g_X^2w^2-v_{\rm ew}^2\left[c_\epsilon^2(g^2+g'^2)-s_\epsilon^2g'^2\right]}\,.
\ee
The mass squared term of neutral vector bosons is transferred to
\be
U_{\rm kin}^\dagger\begin{pmatrix}g'^2v_{\rm ew}^2/4 & -gg'v_{\rm ew}^2/4 & 0 \\ -gg'v_{\rm ew}^2/4 & g^2v_{\rm ew}^2/4 & 0 \\ 0 & 0 & g_X^2w^2\end{pmatrix}U_{\rm kin}=\begin{pmatrix}0 & & \\ & m_Z^2 & \\ & & m_X^2\end{pmatrix}\,,
\ee
where the mass eigenvalues are
\be
m_Z=\frac{gv_{\rm ew}}{2c_W}\sqrt{1+s_Wt_\xi t_\epsilon},\quad
m_{\tilde X}=\frac{g_Xw}{c_\epsilon\sqrt{1+s_Wt_\xi t_\epsilon}}\,.
\ee
Given $|\epsilon|\ll1$ and $v_{\rm ew}\ll w$, they are very close to the decoupling limit values $m_Z\approx gv_{\rm ew}/(2c_W)$ and $m_X\approx g_Xw$.

In classically conformal models one has $m_\phi < m_{\tilde X}$ for perturbative values of $g_X$. Consequently, the $\phi$ boson decays to SM particles only through off‑shell $\tilde X$ exchange. The dominant decay channels are as follows.
\begin{enumerate}
\item Four-body process $\phi\to \tilde X^*\tilde X^*\to f\bar ff'\bar f'$. In the contact‑interaction approximation, corresponding to the dimension‑7 operator $\phi(\bar f\gamma^\mu f)(\bar f'\gamma_\mu f')$, and assuming $m_\phi\gg 100$ GeV, the decay width can be parametrized as
\be\label{4body}
\Gamma_4=\sum_{f,f'}\Gamma_{\phi\to f\bar ff'\bar f'}\approx\mP_4\epsilon^4g_X^{10}w\,,
\ee
where $\mP_4\approx1.60\times10^{-13}$ is determined by numerical simulation.

\item Three-body process $\phi\to\tilde X^*Z\to f\bar fZ+W^+W^-Z$. The triple-boson final state is negligible compared with $\phi\to Zf\bar f$ as it gets an additional suppression factor $v_{\rm ew}^2/m_X^2$, thus
\be
\Gamma_3\approx\sum_{f}\Gamma_{\phi\to Zf\bar f}\approx\mP_3\frac{\epsilon^4g_X^6v_{\rm ew}^2}{w}\,,
\ee
with $\mP_3\approx2.41\times10^{-10}$ being determined by numerical simulation.

\item Two-body process $\phi\to ZZ$, which can be analytically evaluated as
\be
\Gamma_2=\Gamma_{\phi\to ZZ}\approx\frac{g_X^2w}{2\sqrt{6}}\left(1+\frac{9g_X^8w^4}{256\pi^4m_Z^4}\right)\left(\frac{\epsilon\sqrt{g^2+g'^2}g'v_{\rm ew}^2}{4g_X^2w^2-(g^2+g'^2)v_{\rm ew}^2}\right)^4
\approx\frac{9\epsilon^4 g'^4g_X^2v_{\rm ew}^4}{8192\sqrt{6}\pi^4w^3}\equiv\mP_2\frac{\epsilon^4g_X^2v_{\rm ew}^4}{w^3}\,,
\ee
where $\mP_2\approx7.57\times10^{-8}$.
\end{enumerate}
The total decay width is then $\Gamma_\phi=\Gamma_4+\Gamma_3+\Gamma_2$.

In the early Universe, \Eq{V1} in the main text receives thermal corrections, and the finite-temperature effective potential reads
\be\label{VT}
V_T(\phi,T)=V_1(\phi)+\frac{3T^4}{2\pi^2}J_B\left(\frac{g_X^2\phi^2}{T^2}\right)
-\frac{T}{12\pi}g_X^3\left[\left(\phi^2+\frac{T^2}{3}\right)^{3/2}-\phi^3\right]\,,
\ee
where the thermal integral is
\be
J_B(y)=\int_0^\infty x^2\d x\log\left(1-e^{-\sqrt{x^2+y}}\right)\,.
\ee
At high temperatures, the symmetry is restored; as the Universe cools, a potential barrier develops, making the $U(1)_X$-breaking process a supercooled FOPT~\cite{YaserAyazi:2019caf,Mohamadnejad:2019vzg,Khoze:2022nyt,Frandsen:2022klh,Liu:2024fly}. The FOPT dynamics can be calculated via the standard finite-temperature field theory~\cite{Quiros:1999jp}. For $g_X = 0.62$ and $m_X = 10^6$ GeV, the FOPT occurs at a low temperature $T_* \approx 387$ GeV with $\beta/H_n\approx 11.7$. Choosing $\epsilon = 0.25$ yields $\Gamma_\phi/H_* \approx 0.7\times 10^{-3}$, well within the slow-reheating regime, and the resulting EMD allows asteroid-mass PBH with a considerable $f_{\rm pbh}$. Hence, this model naturally contains all necessary ingredients for abundant PBH production. 

\end{document}